\documentclass[aps,prd,reprint,nofootinbib,floatfix]{revtex4-1}
\pdfoutput=1
\usepackage{amssymb,amsmath,graphics,epsfig}
\usepackage{amsmath}
\newcommand{\rem}[1]{ }
\newcommand{\beq}{\begin{equation}}
\newcommand{\eeq}{\end{equation}}
\newcommand{\bea}{\begin{eqnarray}}
\newcommand{\eea}{\end{eqnarray}}

\begin{document}
\title{Plasma Constraints on the Cosmological Abundance of Magnetic Monopoles and the Origin of Cosmic Magnetic Fields}


\author{Mikhail V. Medvedev} 
\altaffiliation{On sabbatical leave from the Department of Physics and Astronomy, University of Kansas, Lawrence, KS 66045}
\author{Abraham Loeb} 
\affiliation{Department of Astronomy, Harvard University, Cambridge, MA 02138}

\begin{abstract}
Existing theoretical and observational constraints on the abundance of magnetic monopoles are limited. Here we demonstrate that an ensemble of monopoles forms a plasma whose properties are well determined and whose collective effects place new tight constraints on the cosmological abundance of monopoles. In particular, the existence of micro-Gauss magnetic fields in galaxy clusters and radio relics implies that the scales of these structures are below the Debye screening length, thus setting an upper limit on the cosmological density parameter of monopoles, $\Omega_M\lesssim3\times10^{-4}$, which precludes them from being the dark matter. Future detection of Gpc-scale coherent magnetic fields could improve this limit by a few orders of magnitude. In addition, we predict the existence of magnetic Langmuir waves and turbulence which may appear on the sky as ``zebra patterns'' of an alternating magnetic field with ${\bf k\cdot B}\not=0$. We also show that magnetic monopole Langmuir turbulence excited near the accretion shock of galaxy clusters may be an efficient mechanism for generating the observed intracluster magnetic fields. 
\end{abstract}

\maketitle
 
\section{Introduction}

Magnetic monopoles are hypothetical particles that carry a net magnetic charge. They have been proposed by \citet{Dirac31} in order to explain the quantization of an electric charge, which is a fundamental experimental fact which, at the time, had no other explanation. Dirac showed that the electric, $e$, and magnetic, $g$, charges must be related by 
\beq
e g = n\hbar c/2,
\eeq
where $n$ is an integer. Thus, the magnetic charge is also quantized, $g = n g_D$, and 
\beq
g_D = \frac{\hbar c}{2e} = \frac{1}{2\alpha}e \approx \frac{137}{2}e
\eeq
is called the `Dirac charge', where $\alpha$ is the fine structure constant. 

Magnetic monopoles are theoretically very attractive because their existence in the Universe would restore the full symmetry of Maxwell's equations:
\bea
\partial_\alpha F^{\alpha\beta}&=&\frac{4\pi}{c} J_e^\beta,
\label{eq1}\\ 
\partial_\alpha \tilde F^{\alpha\beta}&=&\frac{4\pi}{c} J_m^\beta
\label{eq2}
\eea
with the Lorentz force being
\beq
\frac{dp_\alpha}{d\tau}=\left(q_eF_{\alpha\beta}+q_m\tilde F_{\alpha\beta}\right)\frac{v^\beta}{c},
\label{lf}
\eeq
where $F^{\alpha\beta}$ and $\tilde F^{\alpha\beta}=(1/2)\epsilon^{\alpha\beta\gamma\delta}F_{\gamma\delta}$ are the electromagnetic and dual electromagnetic tensors, $J_e=(\rho_e, {\bf j_e})$ and $J_m=(\rho_m, {\bf j_m})$ are the electric and magnetic four-currents, $q_e$ and $q_m$ are electric and magnetic charges, $v$ and $p$ are four-velocity and four-momentum of particles. The action principle for the classical dual electrodynamics has been derived in Ref. \citep{cardoso+96}

\rem{
In vector notations, the equations are
\bea
\nabla\cdot {\bf E} &=& 4\pi \rho_e,
\label{eqE}\\ 
\nabla\cdot {\bf B} &=& 4\pi \rho_m,
\label{eqB}\\ 
-\nabla\times{\bf E} &=& \frac{1}{c}\frac{\partial {\bf B}}{\partial t} + \frac{4\pi}{c} {\bf j}_m,
\label{eqE2}\\
\nabla\times{\bf B} &=& \frac{1}{c}\frac{\partial {\bf E}}{\partial t} + \frac{4\pi}{c} {\bf j}_e,
\label{eqB2}\\
\frac{d {\bf p}}{dt} &=& q_e\left({\bf E} + \frac{1}{c}{\bf v\times B}\right) + q_m\left({\bf B} - \frac{1}{c}{\bf v\times E}\right). ~
\label{eqF}
\eea
}

This system of equations admits symmetry under the duality transformation:
\bea
{{J_e}\choose{J_m}}
=\left(
\begin{array}{cc}
\cos\theta & -\sin\theta\\
\sin\theta & \cos\theta
\end{array}
\right)
{{J_e'}\choose{J_m'}},
\label{dual1}\\
{{\bf E}\choose{\bf B}}
=\left(
\begin{array}{cc}
\cos\theta & -\sin\theta\\
\sin\theta & \cos\theta
\end{array}
\right)
{{\bf E'}\choose{\bf B'}},
\label{dual2}
\eea
for an arbitrary rotation angle $\theta$. Thus, one cannot uniquely assign an electric charge or a magnetic charge (or a mix thereof) to a particle, as they merely become a matter of convention. For example, the roles of the electric and magnetic charges swap upon a rotations of $\theta=\pi/2$.

Furthermore, \citet{tHooft74} and \citet{Polyakov74} discovered the {\em necessity} of magnetic monopoles in Grand Unification Theories (GUT) which unify strong and electroweak interactions. Electric charge in these theories is naturally quantized and the magnetic monopole thus appear almost unavoidably as a topological defect in spontaneous symmetry breaking below the GUT energy scale, with a mass $m_M\sim 10^{17}$~GeV. Larger monopole masses are expected if gravity is involved in a GUT scheme and smaller masses are predicted in theories involving some intermediate scale between the GUT and electroweak energy scales. Magnetic monopoles of the lowest mass (if there are more than one type) must be a stable particle because magnetic charge is conserved. For more details on the theory and observational predictions of magnetic monopoles, see e.g., reviews by \citet{Preskill84} and \citet{PS15}.

Magnetic monopoles are believed to be produced during a phase transition at the GUT energy scale via the Kibble mechanism \citep{kibble80}. Above the critical GUT temperature, $T_c\sim10^{15}$~GeV the symmetry is restored and no monopoles are present. The monopoles appear as topological defects of a scalar field at $T<T_c$. Their abundance is thus set by the correlation length of the scalar field at $T\sim T_c$. Causality limits this length to the horizon scale at that epoch. For an adiabatically expanding Universe, the relic monopole abundance at the present epoch is estimated to be \citep{ZKh78,Preskill79}:
\beq
\Omega_M h^2\simeq 10^{15} \left({T_c}/{10^{15}~\textrm{GeV}}\right)^3 m_{17},
\label{omegam}
\eeq
where $m_{17}=m_M/(10^{17}\textrm{ GeV})$,  $m_M\sim T_c/\alpha$ is the monopole mass, $h = H/(100~\textrm{km s}^{-1}~\textrm{Mpc}^{-1})$ is the normalized Hubble constant and $\Omega_M=\bar\rho_M/\rho_c$ is the density parameter, i.e., the ratio of the average monopole density in the Universe to the critical density $\rho_c=3H/8\pi G\simeq10^{-5}$~GeV~cm$^{-3}$, with $G$ being Newton's gravitational constant. This implies the number density of monopoles would be comparable to that of baryons, thus over-closing the Universe due to their much higher mass, which is impossible. The most attractive solution to this `monopole problem' is inflation, which can dilute the primordial monopole density by a factor of $\sim e^N\sim10^{26}$, where $N\sim60$ is the minimal number of inflation $e$-folds.

Despite extensive searches, magnetic monopoles have never been observed with confidence. The searches include collider experiments, such as MODAL, TRISTAN, PETRA, CDF, D0, HERA, and cosmic ray observatories, such as MACRO, Baikal, Baksan-2, Soudan-2, Ohya,  KGF, AMANDA, ANTARES, IceCube; see the review \citep{PS15} and comprehensive bibliographies \citep{bib1,bib2} for details. Experimental upper limits on the flux of monopoles at Earth are, approximately,
\beq
F_M\lesssim
\left\{
\begin{array}{ll}
10^{-16}~ \textrm{cm}^{-2}~\textrm{s}^{-1}~\textrm{sr}^{-1} & \textrm{ for } v/c \lesssim 0.8,\\ 
3\times 10^{-18}~ \textrm{cm}^{-2}~\textrm{s}^{-1}~\textrm{sr}^{-1} & \textrm{ for } v/c \gtrsim 0.8.
\end{array}
\right.
\label{fluxes}
\eeq
The non-relativistic upper limit is mostly set by the dedicated search with MACRO experiment at Gran Sasso \citep{macro} and the relativistic upper limit is set by IceCube cosmic ray detector in Antarctica \citep{icecube}.

These upper limits are consistent with a theoretical constraint known as the `Parker limit', $F_M<10^{-16} \textrm{ cm}^{-2}~\textrm{s}^{-1}~\textrm{sr}^{-1}$, based on the survival of Galactic magnetic fields \citep{parker70}. Indeed, the work done by the magnetic fields in accelerating monopoles must be replenished by the Galactic dynamo action, thus ${\bf j_m\cdot B}\lesssim (B^2/8\pi)\tau_{dynamo}^{-1}$, where ${\bf j_m}=g_Dn_M{\bf v}$ is the monopole current, $n_M$ is the monopole number density and $\tau_{dynamo}\sim 10^8$~yrs is the typical galactic dynamo timescale. An improved `extended Parker limit' \citep{adams+93} follows from the survival of protogalactic seed fields, yielding
\beq
F_M<10^{-16}m_{17} \textrm{ cm}^{-2}~\textrm{s}^{-1}~\textrm{sr}^{-1}.
\label{e-parker-limit}
\eeq
Monopoles in a cosmological context have also been considered and their energy density in the universe has been derived in Ref. \citep{long+15}.

\section{Monopole plasma}

At the current epoch, the Universe is filled with a fully ionized gas -- plasma. Because of long-range electromagnetic interactions between electrons, protons, and other ions, plasmas support collective instabilities and waves, which are plasma normal modes, such as Alfven and Langmuir waves. Thus, charged particle motions drastically differ from single-particle dynamics in electromagnetic fields. If magnetic monopoles exist and their abundance is large enough, then the symmetry of Maxwell's equations, Eqs. (\ref{eq1})--(\ref{lf}), dictates that the monopole dynamics should exhibit collective motions as well. In this section, we discuss the properties of such  a `{\em magnetic monopole plasma}' and the conditions for such a plasma description to be valid. 

First, we assume that the Universe is {\em magnetically neutral}, that is the amounts of positive and negative magnetic charges are equal so that the net magnetic charge vanishes. This is a convenient `symmetry assumption', though it may be violated and if so, there will be some net (but very weak) magnetic field. Second, it is also likely that the masses of the positive and negative magnetic monopoles are equal, so for simplicity, we assume that as well. Third, we take into account the presence of the ionized gas (i.e., normal plasma) in the Universe. Its dynamics is much faster and, hence, decoupled from that of the monopole plasma since the monopoles are many orders of magnitude more massive than the electrons and ions, whereas their charge is larger by only two orders of magnitude. The role of the ionized gas is crucial, though, because it screens out electric fields and establishes {\em quasi-neutrality}: no large-scale electric fields are present\footnote{We neglect motional electric fields in astrophysical setups where $E\sim (v/c) B \ll B$.}. 

In summary, we assure that the monopole plasma is: (i) `magnetically neutral' (no net charge), (ii) made of particles with the same $|g|/m_M$-ratio, and (iii) has vanishing electric fields. By the duality, given by Eqs. (\ref{dual1}), (\ref{dual2}), this system is very similar to the simplest plasma known: the {\em collisionless unmagnetized electron-position plasma} whose properties are very well studied. For instance, it supports propagation of electromagnetic waves and of longitudinal Langmuir (electrostatic) waves which can be Landau damped. This analogy allows us to proceed with quantitative calculations.

The monopoles should have some nonzero ``thermal'' (random) velocity, $v_{th}$, because they are accelerated by magnetic fields in the same way electric charges are accelerated by electric fields. The kinetic energy gained is $(\gamma-1) m_M c^2=g B l$, where $l$ is the path length and $\gamma$ is the Lorentz factor. The largest systems with magnetic field observed so far are galaxy clusters. The typical intracluster medium magnetic fields have an amplitude of a few micro-Gauss with coherence lengths of tens of kiloparsecs, within the Mpc-cluster scale \citep{vogt+05,bonafede+10,vacca+10,govoni+17,bonafede+13}. A monopole moving through $N_c\sim L/l$ independent patches of coherent $B$-field attains the Lorentz factor, $\gamma$, such that 
\bea
\gamma-1 &=&\frac{g B l\sqrt{N_c}}{m_Mc^2} = \frac{g B \sqrt{ l L}}{m_Mc^2}\nonumber\\
&\simeq& 5.7\times10^{-5}n B_{-6}(l_{-2} L_0)^{1/2} m_{17}^{-1},
\label{gamma}
\eea
where $n$ is an integer, $B_{-6}=B/(10^{-6}\textrm{ Gauss})$ is the typical intracluster field strength, $l_{-2}=l/(10^{-2}\textrm{ Mpc})$ is the field coherence length, $L_0=L/(1\textrm{ Mpc})$ is the size of the magnetized region. Henceforth, we assume the Dirac charge $g=g_D$, so that $n=1$, for simplicity. The general case can easily be restored.

This yields the characteristic thermal velocity to be
\beq
v_{th}/c\simeq
\left\{
\begin{array}{ll}
1,  & \textrm{ if } m_M \lesssim 10^{13} \textrm{ GeV},\\ 
10^{-2}  m_{17}^{-1/2},  & \textrm{ if } m_M \gtrsim 10^{13} \textrm{ GeV}.
\end{array}
\label{vth}
\right.
\eeq
This estimate is about an order of magnitude larger than the previously derived, ($v_{th}/c)\sim10^{-3} m_{17}^{-1/2}$, based on the Galactic magnetic fields, yet it is rather conservative. Indeed, the largest magnetic structures known are {\em radio relics} \citep{bonafede+13,kierdorf+17}. They extend over the distances $L\sim 2$~Mpc, have magnetic fields of strength $B_{-6}\sim 3$ with the coherence length $l\sim L$, based on the lack of substantial variation of polarization of the radio emission. These values yield almost an order of magnitude larger velocity. Since radio relics are rare, however, we do not expect them to contribute much to the energization of the entire cosmic monopole plasma, hence the estimate (\ref{vth}) is adopted hereafter. 

The total number density of monopoles is estimated to be
\beq
n_M=n_+ + n_- 
= \frac{\Omega_M \rho_c \Delta}{\gamma m_M}\simeq (10^{-22}\textrm{cm}^{-3})\,\Omega_M h^2m_{17}^{-1}\Delta,
\label{n}
\eeq
where $n_+$ and $n_-$ are the local densities of positive and negative monopoles and $\Delta=\rho_M/\bar\rho_M$ is the overdensity. Hereafter, our numerical estimates assume non-relativistic monopoles, $\gamma\simeq1$, unless stated otherwise. This density corresponds to the mean distance between the particles of a thousand kilometers or more. 

Unless the monopoles are very massive, they are distributed nearly uniformly, hence $\Delta\simeq1$. However, the current random velocities of particles with $m_M\gtrsim10^{17}$~GeV are comparable to or below the escape velocities from large galaxy clusters, $v_{esc}\sim1000$~km~s$^{-1}$. Thus, such monopoles can be gravitationally trapped with their density being greatly enhanced. For instance, assuming that the monopole density follows the dark matter density for $v_{th}\ll v_{esc}$ as described by the NFW profile \citep{nfw}, the density at the scale radius, $r_s$ (where the velocity dispersion is approximately maximal) is $\rho_s\sim \rho_{vir}(r_{vir}/r_s)^3\sim \rho_{vir}c_*^3$, where $r_{vir}$ and $\rho_{vir}$ are the virial radius and the density at the virial radius and $c_*=r_{vir}/r_s$ is the concentration parameter of the NFW profile. In turn, the dark matter overdensity at the virial radius is typically $\sim 50$. For a typical galaxy cluster, $c_*\sim6$, it yields the monopole overdensity of order $6^3\times50\sim10^4$. Thus, 
\beq
\Delta\sim
\left\{
\begin{array}{ll}
1,  & \textrm{ if } v_{th} \gg 1000  \textrm{ km s}^{-1},\\ 
10^4,  & \textrm{ if } v_{th} \ll 1000  \textrm{ km s}^{-1}.
\end{array}
\label{delta}
\right.
\eeq

Collective plasma excitations have a characteristic frequency -- the plasma frequency -- which in the case of a monopole plasma becomes:
\beq
\omega_{p,M}=\left(\frac{4\pi g_D^2 n_M}{\gamma m_M}\right)^{1/2}
\simeq (3\times 10^{-15}\textrm{s}^{-1})\,(\Omega_M h^2\Delta)^{1/2} m_{17}^{-1}.
\label{omegap}
\eeq
Such excitations can be called, by analogy with normal plasmas, the magnetic Langmuir waves. They have the dispersion relation
\beq
\omega^2=\omega_{p,M}^2+3k^2 v_{th}^2,
\label{langmuir}
\eeq
where $k$ is the wave number, $v_{th}^2=k_B T/m_M$, $T$ is the temperature and $k_B$ is the Boltzmann constant. These waves are caused by charge separation and inertia, and are longitudinal, ${\bf k || \tilde B}$, because the perturbed field is no longer divergence-free: $\nabla\cdot\tilde {\bf B}=4\pi g_D(n_+ - n_-)$. 

The magnetic Langmuir wave, by analogy with the normal one, should experience collisionless (Landau) damping, which is particularly strong when the wave phase velocity is comparable to the thermal velocity, $v_{ph}=\omega/k\simeq v_{th}$. The Landau damping rate, defined as the imaginary part of a complex frequency, is given by:

\beq
\Gamma_{Landau}\simeq \frac{\pi\omega_{p,M}}{2k^2n_M}\left.\frac{df_0(v)}{dv}\right|_{v=\omega/k},
\label{landau}
\eeq
where $f_0(v)$ is the unperturbed distribution function of monopoles and $\int f_0\,dv=n_M$. Note that the wave fields are $\propto\exp(i{\bf k\cdot x}-i\omega t+\Gamma t)$. Hence damping occurs when $df_0/dv$ is negative, as it is for the Maxwellian distribution function, for example. 

An electromagnetic wave is another normal mode in such a plasma, with a very similar dispersion relation, $\omega^2=\omega_{p,M}^2+k^2 c^2$. The characteristic spatial scale associated with this eigenmode is the skin length:
\beq
d=c/\omega_{p,M} \simeq (10^{25}\textrm{cm})\,(\Omega_M h^2\Delta)^{-1/2}  m_{17}.
\eeq
Since $\omega_{p,M}$ is extremely small, the electromagnetic wave propagation is not significantly affected. A low-frequency acoustic mode, with the dispersion relation $\omega^2=v_s^2k^2$, can also exist in the monopole plasma, where $v_s^2=\hat\gamma v_{th}^2$ is the sound speed and the effective adiabatic index $\hat\gamma\gtrsim1$. Since $v_s\sim v_{th}$, this mode is efficiently damped by Landau damping, as described by Eq. (\ref{landau}). 

The Debye length in normal plasmas characterizes screening of electric fields. Similarly, the magnetic Debye length determines the scale above which the plasma is magnetically quasi-neutral,
\beq
\lambda_D=\frac{v_{th}}{\omega_{p,M}}
\simeq (10^{23}\textrm{cm})\,(\Omega_M h^2\Delta)^{-1/2}  m_{17}^{1/2}.
\label{lambdad}
\eeq
At large distances from the source, $r>\lambda_D$, the field is exponentially suppressed, $\propto e^{-r/\lambda_D}$. The time-scale on which quasineutrality is established is fast: $\tau_{qn}\sim\lambda_D/v_{th}\sim\omega_{p,M}$. 

So far, we have assumed that the monopole distribution is dense enough to be treated as a plasma. For this to hold true, the plasma parameter -- the total number of particles within a Debye sphere -- must be much greater than unity. Indeed,
\beq
N_D=\frac{4\pi}{3}\lambda_D^3 n_M
\simeq 6\times10^{47}\,(\Omega_M h^2\Delta)^{-1/2} m_{17}^{1/2}.
\eeq
Thus, unless the abundance $\Omega_M\ll10^{-90}$ or so, the plasma condition is safely satisfied: $N_D\gg1$. In fact, the monopole plasma, if it exists, is the best plasma in the Universe. 

Finally, the characteristic time between particle collisions is $\tau=1/\nu$, where the collision frequency is
\beq
\nu=\frac{4\pi g_D^4 n_M\ln\Lambda}{m_M^2 v_{th}^3}
\simeq (10^{-63}\textrm{ s})\, \Omega_M h^2\Delta m_{17}^{-3/2} \ln\Lambda,
\eeq
where $\ln\Lambda=\ln(\lambda_D/r_0)\sim120$ is the magnetic Coulomb logarithm and $r_0\sim g_D/m_M v_{th}^2$ is the distance of the closest approach. Thus, the plasma is highly collisionless with the collision time being $\tau\sim 10^{43} t_H$, where $t_H\sim4\times10^{17}$~s is the Hubble time.

\section{Abundance constraints}

The existence of astrophysically strong, micro-Gauss magnetic fields on Mpc scales places a tight constraint on the monopole abundance. The monopole plasma should screen out magnetic fields on scales greater than the magnetic Debye length to make it quasi-neutral. The largest scale fields are observed in clusters and radio relics \citep{bonafede+13,kierdorf+17}, whose scale is $L\simeq 2$~Mpc, thus,
\beq
\lambda_D > L.
\label{constr}
\eeq
This sets the upper limit on the monopole abundance:
\beq
\Omega_M h^2 < 10^{-3}\Delta^{-1}m_{17}.
\eeq
Note that this limit rules out entirely the possibility that monopoles (of sub-Planckian mass) make the dark matter, even for $\Delta=1$, i.e., without taking gravitational trapping into account. 

Furthermore, the overdensity $\Delta$ depends on $v_{th}$ which is a function of $m_M$, see Eqns. (\ref{vth}), (\ref{delta}). A detailed exploration how gravitational trapping of monopoles occurs in dark matter halos goes beyond the scope of this paper. Instead, we adopt here a simple function that smoothly interpolates between the two limiting cases in Eq. (\ref{delta}) as follows:
\beq
\Delta = 1 + 10^4\tanh^4\left[(3\times10^7\textrm{ cm s}^{-1})/v_{th}\right].
\label{delta-interp}
\eeq
It yields $\Delta\sim10^4$ for $v_{th}\lesssim300$~km~s$^{-1}$, $\Delta\sim10^2$ at $v_{th}\sim1000$~km~s$^{-1}$ and $\Delta\sim1$ for $v_{th}\gtrsim3000$~km~s$^{-1}$. 

\begin{figure}
\vskip1em
\centering
\includegraphics[scale = 0.65]{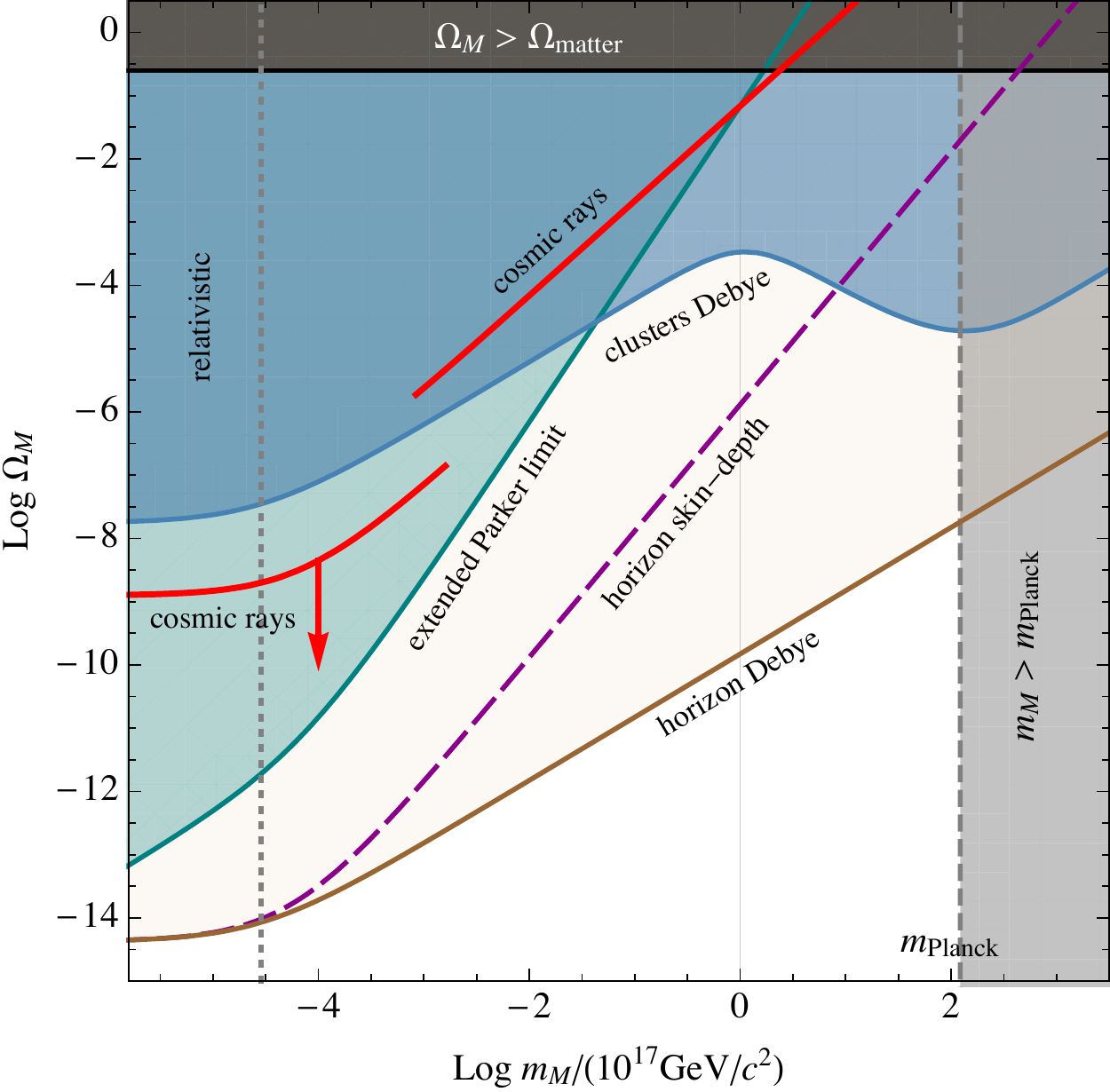}  
\caption{Monopole mass-abundance diagram. The blue shaded region labeled `clusters Debye' is excluded based on the plasma constraint, i.e., that the magnetic Debye scale must exceed the galaxy cluster scale. The extended Parker exclusion region is shown as a shaded green area. The black region at the top is excluded because the monopole mass density exceeds that of matter. The grey shaded area on the right is the super-Planckian mass region. Monopoles are relativistic to the left of the vertical dotted line. The observational upper limits are shown as a disjoint red curve with the down arrow. The purple dashed and solid brown curves show conditions when the plasma skin depth and the Debye scale are comparable to the horizon scale of the present day Universe, respectively. 
}
\label{fig} 
\end{figure}

Figure \ref{fig} shows the abundance constraint represented by Eq. (\ref{constr}): the blue shaded region is the exclusion region. Since this constraint follows from the fact that the Debye length exceeds the galaxy cluster scale, it is labeled as ``clusters Debye". Note that this region is computed using exact Eqs. (\ref{lambdad}), (\ref{omegap}), (\ref{n}), (\ref{gamma}) and the interpolating function (\ref{delta-interp}) for $\Delta$. The exclusion region from the extended Parker limit, Eq. (\ref{e-parker-limit}), is also shown as the green shaded area. The black horizontal region at the top is excluded because $\Omega_M$ exceeds that of the matter, which is impossible. The grey band on the right corresponds to monopole masses exceeding the Planck mass $m_{Planck}=(\hbar c/G)^{1/2}\simeq1.22\times 10^{19}$~GeV. The vertical dotted line marks where, $v_{th}\sim c$, i.e., the magnetic monopoles are relativistic to the left of this line. The observational upper limits in Eq. (\ref{fluxes}) are shown as a red broken curve with the downward arrow. 

Figure \ref{fig} implies that the Debye limit strengthens the abundance constraint by several orders of magnitude in the most interesting region of monopole masses. Moreover, there is an absolute upper limit on the abundance:
\beq
\Omega_M\lesssim3\times10^{-4}.
\label{abslimit}
\eeq
That is, the contribution of monopoles to the overall mass budget of the Universe is highly subdominant, irrespective of their mass. In fact, the plasma constraint is not strongly mass dependent. The $\Omega_M$ upper limit is within the range of $10^{-4}-10^{-5}$ for the range of masses $10^{15}\lesssim m_M\lesssim10^{19}$~GeV. This is largely because of the gravitational clustering of monopoles in large dark matter halos of galaxy clusters, which becomes important at masses $m_M\gtrsim10^{17}$~GeV. For masses $m_M\lesssim10^{15}$~GeV, the extended Parker limit is more stringent. The above absolute limit assumes the estimate of $v_{th}$ from Eq. (\ref{vth}) and hence slightly depends on it, being less stringent for larger $v_{th}$. 

From the absolute $\Omega_M$-limit, Eq. (\ref{abslimit}), one obtains the number density at Earth using Eq. (\ref{n}) with $h^2\Delta\sim1$ to be $n_M\lesssim10^{-26}$~cm$^{-3}$. This corresponds to the monopole flux upper limit:
\beq
F_M\lesssim3\times 10^{-19} m_{17}^{-3/2} \textrm{ cm}^{-2}~\textrm{s}^{-1}~\textrm{sr}^{-1},
\label{flimit}
\eeq
which is significantly tighter than the previous limit in Eq. (\ref{fluxes}). Note that the possible detection of the fields of strength $\sim 10^{-14}$ micro-Gauss at scales of 10 Mpc \citep{long+15} does not improve the above upper limit because the overdensities at these scales are of order unity. 

Figure \ref{fig} also shows the conditions when two characteristic scales, the plasma skin depth and the magnetic Debye length, are comparable to the present day horizon scale. We assume $\Delta=1$ for these curves because no gravitational trapping occurs on this scale. The skin depth represents a characteristic scale of various electromagnetic plasma phenomena. The horizon-scale skin depth (shown by magenta dashed curve) shows the conditions when these scales are comparable. Below the line, the skin depth exceeds the observed size of the Universe. 

Even more interesting is the horizon-scale magnetic Debye length (shown in Figure \ref{fig} as a solid brown line) because the Debye length represents a characteristic scale below which plasma provides no shielding effect. Obviously, below this line, the magnetic Debye length is larger than the size of the Universe, i.e., no shielding of magnetic fields at all scales is possible. Thus, we predict that if future observations will detect fields coherent on Gpc scales, this would exclude the region above the curve and thus further limit the monopole abundance by orders of magnitude, to the region $\Omega_M<10^{-8}$. This value is, coincidentally, close to the theoretically predicted abundance given by Eq. (\ref{omegam}) after diluted by a factor $\sim e^{60}$ by inflation: $\Omega_M\sim10^{-11}m_{17}^4$. If inflation proceeds longer, the dilution of monopoles  by $\sim e^N$ ($N$ being the number of $e$-folds) will proportionally reduce their current $\Omega_M$ well below $10^{-11}$, where the plasma collective effects are  harder to observe.

\section{Magnetic Langmuir waves}

If $\lambda_D$ is smaller than the horizon scale, we predict an interesting new phenomenon -- the magnetic Langmuir wave. Its dispersion relation is given by Eq. (\ref{langmuir}). Like its conventional electrostatic counterpart, this wave is caused by charge separation and inertia. It can be excited by time-dependent by dynamic magnetic fields in galaxy clusters, in jets and outflows from galaxies, as well as by monopole plasma instabilities driven by flow inhomogeneities during the formation of the large-scale structure, e.g., assembly of dark matter sheets and filaments, and mergers of galactic and cluster halos. 

Observationally, such a wave may be discerned via its ``zebra pattern'' of an alternating magnetic field with a characteristic wavelength $\lambda_D\sim\lambda_{wave}\sim 1/|{\bf k}|$, where ${\bf k}$ is the wave vector. This pattern can be detected with Faraday rotation and synchrotron emission by energetic electrons. The ``smoking gun'' signature of the wave would be the alignment of ${\bf k}$ and ${\bf B}$ vectors on the sky. This is because a Langmuir wave is longitudinal, ${\bf k\cdot B}\not=0$, i.e., the $B$-field is manifestly {\em non-divergence-free}. Without monopoles, $\nabla\cdot {\bf B}=0$ implies ${\bf k\cdot B}=0$ identically, i.e., the field is aligned with the interfaces, as it is seen in tangential discontinuities, for example. Whether the pattern has the ${\bf k\cdot B}\not=0$ signature, can be inferred from radiation polarization measurements of the field orientation. 

The amplitude of the wave can readily be estimated from $\nabla\cdot{\bf B}=4\pi g_D(n_+ -n_-)$, yielding
\bea
B&\sim&4\pi g_D\delta n_M\lambda_{wave}
\\
&\sim&(100\textrm{ Gauss})\left(\frac{\Omega_M}{3\times10^{-4}}\right)^{1/2} \Delta^{1/2} m_{17}^{-1/2}\,\frac{\delta n_M}{n_M},
\nonumber
\eea
where $\delta n_M/n_M = (n_+ -n_-)/n_M$ is the dimensionless density perturbation due to the magnetic charge separation, and we assumed that $\lambda_{wave}\sim\lambda_D$. 

Together with the observational upper limit of $B<10^{-15}$~Gauss at scales $>100$~Mpc, this implies very small density perturbations of order ${\delta n_M}/{n_M}\lesssim10^{-17}$ if $\Omega_M\sim10^{-4}$. Obviously, such waves are in the linear regime amenable to theoretical studies. Of course, the ``zebra pattern'' would only be seen if a single wave is excited. Quite often, an entire spectrum of waves is present in plasmas. These waves would be seen as just standard turbulent $B$-fields with some spectral distribution. The turbulence will rather look like  a ``leopard spots'' pattern similar to that of the magnetohydrodynamic (MHD) turbulence. Projection effects can smear the pattern, though. Observational demonstration of ${\bf k\cdot B}\not=0$ in this turbulence may be difficult.

\section{Origin of cosmological fields}

Interestingly, no magnetic fields on scales larger than a few Mpc have so far been reliably observed. The upper limit on the void fields (on scales of a Gigaparsec) is approximately $<10^{-15}$~Gauss \citep{voidB}, a few orders of magnitude smaller than the nano-Gauss fields needed to explain the intracluster magnetic fields by gas compression in the accretion process, without additional field amplification \citep{dt08}. Remarkably, this fact is consistent with being due to the shielding of the fields at scales $\gg \lambda_D$ if $\lambda_D\sim1$~Mpc. Of course, the absence of the magnetic field sources is another possible explanation. 

The absence of $B$-field sources outside clusters brings up a question about the origin of magnetic fields {\em in clusters}. If $\Omega_M\sim3\times10^{-4}$, the observed fields may be created by turbulence in the monopole plasma during structure formation and accretion and possibly further amplified by compression and MHD turbulence in the intracluster ionized gas.

As the large scale structure forms, dark matter, baryonic matter and monopoles nearly follow each other until shell crossing. At this moment, dark matter forms multiple streams because it is collisionless and is non-interacting via long-range forces other than gravity. In contrast, ionized gas cannot form a multi-stream state on large scales because plasma instabilities generate strong electromagnetic fields \citep{sagdeev-text}. These fields exist on small, kinetic scales and thus act as {\em effective collisions} that scatter particles over pitch angle to isotropize their motion and thermalize their distribution function. This mechanism establishes collisionless accretion shocks. Without such electromagnetic turbulence, shocks would not exist because the Coulomb mean free path often exceeds the system size by orders of magnitude.  There are many mechanisms that generate such electromagnetic fields, depending on specific conditions in the medium. For example, in non-magnetized plasmas, Weibel instability is the primary process \citep{ml99,msk06}. 

By symmetry, the monopole plasma will behave similarly. In the presence of the ionized gas, however, electric fields cannot be efficiently generated by the monopole plasma instabilities because they are short-circuited by the currents in the ionized gas. Thus, only {\em magnetic Langmuir turbulence} can be produced. The resulting effect is a very efficient beam-plasma two-stream instability, which is essentially inverse Landau damping. If the streams have sufficiently large initial thermal velocities, the growth rate, $\Gamma$, is given by Eq. (\ref{landau}), which is valid if $v_{th}\gg \Gamma/k$. In the opposite case of cold plasmas, $v_{th}\ll \Gamma/k$, the entire beam is in Landau resonance and the growth rate for $k<\omega_{p,M}/u$ is 
\beq
\Gamma=k u (n_1/n_0)^{1/2} (k^2 u^2/\omega_{p,M}^2-1)^{-1/2},
\eeq
where $n_0$ and $n_1$ are the densities of the bulk plasma and the beam and $u$ is the velocity of the beam. For the accretion shock, $n_0$ and $n_1$ can be treated as the downstream and upstream densities (hence $n_1\sim n_0$) and $u$ is approximately the upstream velocity in the shock frame. The maximum growth rate and the corresponding wave number are 
\beq
\Gamma_{max}=\omega_{p,M}\frac{3^{1/2}}{2^{4/3}}\left(\frac{n_1}{n_0}\right)^{1/3}\sim \omega_{p,M}, \quad
k_{max}=\omega_{p,M}/u.
\eeq

The amplitude of the generated fields must be large in order to efficiently scatter and thermalize particles in otherwise collisionless plasma. Thus, the energy density in the magnetic Langmuir turbulence should be comparable to the kinetic energy density of the flow, $B^2/8\pi\simeq m_M n_M u^2/2$. This yields
\bea
B&=&\left(4\pi m_M n_M u^2\right)^{1/2} \\
&\simeq&(3\times10^{-7}\textrm{ Gauss})\left(\frac{\Omega_M}{3\times10^{-4}} \frac{\Delta}{200}\right)^{1/2} 
 \left(\frac{u}{10^3\textrm{ km s}^{-1}}\right),
\nonumber
\eea
where we assumed the typical overdensity $\Delta\sim200$ at the location of the accretion shock near the virial radius, and accounting for the shock compression.

These fields are generated on the time-scale of $\tau\sim1/\Gamma_{max}\sim1/\omega_{p,M}\sim6\times10^7\, m_{17}$~yrs and have a characteristic scale of $\lambda_B\sim1/k_{max}\sim u/\omega_{p,M}\sim60\, m_{17}$~kpc. Note that once these magnetic fields are created, they may be maintained and further amplified by currents and turbulence in the ionized gas in the cluster, because its dynamics should dominate over that of the monopoles, since $\Omega_{matter}\gg\Omega_M$. Indeed,  these fields can be the seed fields which may be amplified by compression during gas accretion toward the center of a cluster, as well as by magnetic turbulent dynamo in clusters and galaxies. 

The earliest epoch when cosmological magnetic fields can be generated by this mechanism is the first shell crossing when Zel'dovich pancakes start to form. Assuming the characteristic redshift of $z\sim20$, the monopole density should be a factor of $(1+z)^3$ larger than that at present, Eq. (\ref{n}), with $\Delta\sim1$. Also assuming $u\sim1\textrm{ km s}^{-1}$ as a typical speed at that epoch, we estimate that the Zel'dovich pancakes should be magnetized with the field of magnitude $B\sim2\times10^{-9}$~Gauss and characteristic scale of $\lambda_B\sim1$~kpc. It may be very difficult to observe these fields (unless the pancake is seen nearly edge-on) because of the small filling factor of the pancakes. If some fast radio bursts originate at such high redshifts \citep{fl16}, they can be used to detect these fields by Faraday rotation.

\section{Conclusions}

In this paper we demonstrated that if magnetic monopoles exist, then they would form a plasma, whose properties are very similar to those of a collisionless electron-positron plasma without magnetic fields. The plasma collective effects place a strong constraint on the monopole abundance. Particularly, the existence of micro-Gauss magnetic fields in galaxy clusters and radio relics implies that the Debye scale is larger than a Mpc. This sets a universal upper limit on the monopole abundance and flux, as stated in Eqs. (\ref{abslimit}), (\ref{flimit}). We predict the existence of magnetic Langmuir waves which may appear on the sky as ``zebra patterns'' of an alternating magnetic field with the wavelength being of order the Debye length. We also predict that if coherent magnetic fields are observed on scales comparable to the horizon scale, this will further limit the monopole abundance by a few orders of magnitude. However, if the cosmological density parameter of monopoles, $\Omega_M$, is well below $\sim 10^{-8}$, the Debye length exceeds the horizon size, implying that the monopole plasma is unable to screen any magnetic fields. Finally, we find that the currently observed magnetic fields could be generated by monopole plasma instabilities, which predict fields as strong as 0.3 micro-Gauss at the virial shock of a galaxy cluster. The coherence scale of these fields is estimated to be of order 100~kpc. These fields can be amplified by accretion via compression and MHD turbulence in the ionized intracluster gas. It would be interesting to use numerical simulations to further investigate these processes in the future.

\acknowledgements

MM is grateful to the Institute for Theory and Computation at Harvard University for support and hospitality and acknowledges partial support via grant DE-SC0016368. This work was supported in part by the Black Hole Initiative, which is funded by a grant from the Templeton Foundation.

\end{document}